\documentclass[10pt]{article}
\usepackage[T1]{fontenc}
\usepackage[latin1]{inputenc}
\usepackage{amsmath}

\makeatletter

\newcommand{\LyX}{L\kern-.1667em\lower.25em\hbox{Y}\kern-.125emX\spacefactor1000}

\newcommand{\lyxaddress}[1]{
  \par {\raggedright #1 
  \vspace{1.4em}
  \noindent\par}
}

\makeatother

\begin{document}

\title{From features to expression: High-density oligonucleotide array analysis revisited.}

\author{F\'elix Naef \protect\( ^{1}\protect \), Daniel A. Lim\protect\( ^{2}\protect \),
Nila Patil\protect\( ^{3}\protect \), and Marcelo O. Magnasco\protect\( ^{1*}\protect \)}

\maketitle

\lyxaddress{The Laboratories of \protect\( ^{1}\protect \)Mathematical Physics and \protect\( ^{2}\protect \)Neurogenesis,
The Rockefeller University, New York NY 10021\\
\protect\( ^{3}\protect \)Perlegen Sciences Inc./Affymetrix Inc., Santa Clara,
California 95051 }

\begin{abstract}
One of the most popular tools for large scale gene expression studies are high-density
oligonucleotide (GeneChip$^{\mbox{{\footnotesize ®}}}$) arrays. These currently
have 16-20 small probe cells (``features'') for evaluating the transcript
abundance of each gene. In addition, each probe is accompanied by a mismatched
probe (MM) designed as a control for non-specificity. An algorithm is presented
to compute comparative expression levels from the intensities of the individual
features, based on a statistical study of their distribution. Interestingly,
MM probes need not be included in the analysis. We show that our algorithm improves
significantly upon the current standard and leads to a substantially larger
number of genes brought above the noise floor for further analysis.

.
\end{abstract}
Bioinformatics is based on the existence of vast quantities of information of
unknown significance whose internal relationships are analyzed using statistical
methods. The individual data in these data sets are usually highly inhomogeneous
in quality, with the number of elements increasing rapidly for lower quality
levels. A recurrent problem in the statistical analysis of such data sets is
that while no sophisticated methods are needed to ascertain the meaning of the
few high quality elements, the bulk of the data often lies near the noise floor,
where fairly fancy statistical tools may become necessary. In such circumstances,
seemingly innocuous improvements to data treatment may yield large improvements
to the analysis simply because of the way the data quality is distributed. 

Among the many experimental techniques generating large datasets from biological
experiments today, oligonucleotide hybridization arrays have rapidly become
a popular tool for large scale gene expression screens\cite{Lander,Lockhart}.
Currently, DNA hybridization array techniques aim at obtaining several thousand
low quality measurements of transcript abundance in a single parallel experiment.
From this ``bulk'' data, the goal is to identify groups of genes participating
in a given pathway and hopefully unravel some features of their transcriptional
regulation, to be confirmed by more sensitive and precise methods. 

There are currently two main trends in microarray technology, cDNA bicolor glass
slides \cite{cDNA_slides,cDNA_slides2} and the high-density oligonucleotide
arrays (HDONAs) manufactured by Affymetrix \cite{Affy_paper1, Affy_paper2}.
In the first case, PCR-derived cDNAs from libraries are spotted onto a glass
slide as hybridization probes. In the second, hybridization probes consist of
chemically synthesized \( 25 \)-mer oligonucleotides on a grided array. Under
the best conditions, one would expect a linear relationship between the measured
fluorescence and the concentration of original mRNA. However, the constant of
proportionality is currently strongly dependent on the hybridization sequences.
As a consequence, large scale hybridization experiments do not give quantitative
information on a gene vs. gene fashion for a single preparation, i.e., it is
not possible to infer the ratio of mRNA concentration for actin to tubulin within
a single sample. The meaningful information lies in the ratios of intensities
for the same hybridization sequence taken from different samples. Usually one
thinks of one sample (e.g. 'normal' tissue or unsynchronized cells) as a baseline
to which all other conditions are compared. 

In what follows, we concentrate exclusively on HDONAs. On these, \textit{probe
cells} (or features) are grouped into \textit{probe sets} for a given gene,
a probe set consisting of \( \sim 20 \) (depending on the gene and the chip
series) \textit{probe pairs} (pairs of probe cells). Each pair is designed to
probe a different 25 base sequence (the identity of which is not currently revealed
by Affymetrix) from the gene. To check for non-specific hybridization, each
probe pair consists of a PM (Perfect Match) cell containing the exact sequence
of that gene and a MM (single MisMatch) cell whose central position (the 13th
nucleotide of the 25-mer) has been substituted. Hence, a full probe set consists
of \( \sim 40 \) hybridization probes, and \textit{composite} scores (for intensity
or ratios) must be derived for each gene. The composites usually used are the
ones provided by default by the Affymetrix software. To generate an absolute
intensity measure (Avg Diff), this algorithm subtracts the MM from the PM intensity
for each probe pair (an attempt to correct for the non-specific hybridization
and background), and the obtained differences are then arithmetically averaged
after truncation of the largest and smallest values \cite{Affy_paper1, schadt}.

We propose an improved method for obtaining composite ratios (and intensities)
of transcript abundance between two samples, based on a study of the statistical
distribution of individual cell ratios within each probe set. A few ideas guided
our approach: (i) the experimental protocol is designed such that the hybridization
is kinetically dominated; (ii) data distributed on an exponential scale should
not be averaged algebraically but geometrically. Having these in mind, we show
how the study of raw Affymetrix data (.CEL files) leads to an algorithm whose
essential ingredients are: (i) MM cells are not utilized as controls for non-specificity,
we use them only for the calculation of the background intensity; (ii) outliers
need to be discarded; (iii) averages are taken in log-coordinates. Significant
advantages over the current Affymetrix software include (i) the ability to obtain
reliable scores for a greater proportion of genes (\( +30\% \)), especially
in the mid to low intensity range; (ii) replicate experiments show greater reproducibility
(i.e. tighter scatter plots); (iii) ratio scores for genes probed twice or more
on the microarray show a vastly increased correlation (there are \( \sim 700 \)
such genes on the combined Mu11k A and B mouse chip series). In the remaining,
we demonstrate how this procedure emerges from studying the data sets, and report
evidence for the improvements.

\section*{Raw intensities and background subtraction}

The starting point in our analysis of HDONAs consists of the fluorescent intensities
of all the 25-base probes on the chip, including both the PM and MM cells (.CEL
files). This data has already gone through one processing step by Affymetrix,
namely an average of the pixel intensities (36 pixels per cell for the Mu11k
mouse chip) for each probe cell. Before moving on to consider which probes belong
to which gene, it is instructive to inquire about the reproducibility of the
raw data in replicate experiments. The cigar shaped cloud in Fig. 1(a) shows
such a typical example. In an ideal (noiseless) experiment, the scatter plot
of the replicates should produce a single straight line with unit slope, so
that the broadening of the line in a real experiment reflects the noise. In
HDONAs, this noise has multiple sources, including intrinsic biological and
sample processing variability, hybridization kinetics and thermodynamics, noise
related to the incorporation and amplification of fluorescent dyes, and the
measurement of the fluorescence in the scanning process\cite{schadt}. Despite
all these potential sources, the experimental situation is encouraging as indicated
by the high reproducibility in (a). 

The intensity dependence of the noise envelope is commonly referred to as the
\textit{noise funnel}. In Fig. 1(a), the funnel is only very weakly intensity
dependent. We observe that the onset of the intensities is shifted from zero
to \( \sim 500 \), indicating that cell intensities have an additive background
component. Estimating this background intensity is essential when processing
low intensity data points. In HDONAs, identifying background is a priori a different
problem from its analogue in cDNA spotted arrays. There, one tries to measure
the intensity of regions in between adjacent spots as a measure for the local
background. In contrast, the inter-feature distance in Affymetrix arrays is
too small for a similar measurement and one must estimate the background from
the probe cells themselves. Background is by definition non-specific, and should
therefore not be sensitive to the single base sequence modification in the MM
cells. Consequently, we consider the subset of probe pairs whose PM and MM intensities
differ by less than a given small quantity (PM-MM < \( \epsilon  \)) as representative
of the background. The distributions of either the PM or MM cells obtained in
this manner depend only weakly on \( \epsilon  \), and can be reasonably fitted
to Gaussians from their low-intensity onset up to their maximum (Fig. 1(d)).
We used \( \epsilon =50 \) in units of the .CEL file intensities, but using
\( \epsilon =100 \) leads to changes of the order of only \( \sim 1\% \) for
the mean background \( \langle b\rangle  \) and standard deviation \( \sigma  \).
Fig. 1(b) shows the raw data after background subtraction. The typical broadening
of the \textit{}noise funnel at low intensities end reflects the residual background
(the fact that \( \sigma \neq 0 \)). In contrast, the Affymetrix procedure
estimates \( \langle b\rangle  \) and variance var\( (b) \) from the \( 2\% \)
lowest intensity cells. The mean and variance obtained this way are strongly
dependent on the arbitrary cutoff (\( 2\% \)). Typically, we obtain a \( \langle b\rangle  \)
larger than the value reported by the Affymetrix software \( (\sim +15\%) \),
so that we are left with \( \sim 86\% \) of the features lying above \( \langle b\rangle +2\sigma  \),
rather that \( \sim 93\% \). In addition, our noise funnel is slightly broader.
Nevertheless, our algorithm for composite scores still leads to a significant
noise reduction (cf. Results). 

As a preview, we show in Fig. 1(c) which subset of PM probes are considered
by our algorithm when computing ratio scores between the two samples. It turns
out consistently that noisier cells are automatically discarded, however, not
on the basis of an evaluation of the funnel shape (cf. Probe cell selection).

\section*{Constructing good estimators}

The previous discussion about background intensity distributions raises the
following general issue: what are good estimators for data drawn from an unknown
distribution? The answer involves finding the coordinates in which the distribution
is most well behaved. In the best situation, a distribution is short tailed,
which ensures that moments are not only well defined but are also relevant quantities
for a statistical description of the data set. In a situation of long tailed
distributions (e.g. driven by a large number of outliers in a dataset), the
situation is more complicated. Then, one either needs to establish a model describing
how one should truncate the dataset before calculating averages, or work with
estimators which are more robust to outliers, like percentiles. To formulate
this more precisely, we consider the following problem: suppose we have \( n \)
samples from a positive distribution \( p(x) \), and samples from the scaled
distribution \( \lambda p(x/\lambda ) \). The problem is to find the optimal
estimator for \( \lambda  \). The solution clearly shall depend on \( p \).
If \( p \) is a well behaved distribution, then \( \langle \lambda p(x/\lambda )\rangle /\langle p(x)\rangle  \)
(\( \langle x\rangle =\frac{1}{n}\sum _{i=1}^{n}x_{i} \) denotes the arithmetic
average) is a fine estimator; but it miserably fails if \( p \) is long tailed.
Conversely, the median is a suboptimal estimator in the case of well behaved
distributions, but it has the advantage of being more robust in the long tailed
case.

Datasets from HDONAs do exhibit such long tails, as we show in Fig. 2. The histograms
(Fig. 2A) show the \( \log _{2} \) PM intensity distributions of all probe
sets, each PM cell being normalized by their probe set median. Probe sets are
classified into four windows according to their median magnitude; we have verified
that refined windows do not change the shape of the distributions significantly.
These distributions show that cell intensities vary by factors of \( \sim 2^{5} \)
around their median in all the intensity windows, and that the distributions
are far from Gaussian. Nevertheless, log-coordinates lead to roughly symmetric
distributions, at least up to the last intensity window. We think of these distributions
as the sum of a well behaved component (with converging moments to which the
Central Limit Theorem applies), plus a long tailed part due to outliers. These
need to be identified (cf. Probe cell selection) and discarded. Then, the meaningful
estimators for the truncated data sets consist of arithmetic averages in log-coordinates
(geometric means).

The ordered intensity profiles of individual, randomly picked probe sets are
also shown (Fig. 2B), each of them for duplicate experiments. These emphasize
the reproducibility of the broad intensity profiles.

\section*{The mysterious MM cells}

Before explaining how to discard outliers and compute a ratio score, we explain
why we do not utilize the MM cells for the calculation of composite intensity
and ratio scores. Single mismatch cells seem not to be consistently doing what
they were originally designed for, namely to serve as a control for non-specific
hybridization. Instead, we find that MM cells often act as a pale PM, essentially
binding the same oligonucleotide as the PM do, but on average \( \sim 1.8 \)
times weaker than the PM probes (Fig. 3). Notice that (a) presents the raw cell
intensities, whereas (b) reports the distributions of composite intensities
obtained by considering the PM and MM probe sets as if they were two different
conditions for the same gene. It is somewhat disturbing that in the high intensity
region, the cloud (a) exhibits a valley around the diagonal. This means that
there is a significant number of probes where the target cRNAs bind more specifically
to the MM (in contrast, non-specific hybridization would result in a maximum
on the diagonal). A possible scenario for this matter are polymorphisms between
the mouse specie used to design the probe sequences and that used as the target.
In such instances, the MM may actually act as the effective PM, however, these
should be rare events since the MM sequence is always substituted exactly at
the central position. 

In any case, subtracting the MM from the PM intensities is likely to be misleading,
and we found it favorable to not consider the MM cells any further. As a matter
of fact, it is not entirely surprising that a single base change does not provide
a clear cut discrimination for non-specific hybridization in a process dominate
by kinetics rather that equilibrium thermodynamics.

\section*{Probe cell selection and ratio composites}

We now describe our algorithm for the selection of cells used in the calculation
of ratios. We observed that comparing two identical probe sets hybridized to
two different samples leads to series of pairwise PM cells ratios \( (r_{1},r_{2},\dots ,r_{N}) \)
behaving quite far from an ideal homogeneous situation (all \( r_{i} \) being
the same). Instead, the individual cell ratios often vary over a decade; it
also occurs that some cells indicate an up-regulation whereas others indicate
the opposite (cf. Fig. 4 (c) and (d)). In this situation, a straightforward
linear regression between PM intensities of the two samples is not adequate.
It further happens that high intensity cells saturate in one or both of the
samples, leading to useless (even misleading) cell ratios. Such probe cells
are discarded from our analysis. The saturation thresholds (most likely due
to the photomultiplier) can be read off the .CEL files, by plotting the mean
cell intensities versus the standard deviation of the pixel intensities. Our
purpose here is not to address the question of why such broad ratio distributions
may arise, but rather how to optimize scores for them. In a first step, we order
the series \( (\log r_{1},\log r_{2},\dots ,\log r_{N}) \). Next, we aim at
splitting this set in an interval \( I_{med} \) with optimally narrow range,
and a subset of outliers to be omitted from the ratio score calculation. We
require the median to be a member of \( I_{med} \) and optimize for its left
and right boundaries \( i_{l} \) and \( i_{r} \). \( l=\log r_{i_{r}}-\log r_{i_{l}} \)
denotes the range of \( I_{med} \) and \( L \) the range of the full probe
set. In the absence of knowledge about how the ratios are distributed within
\( I_{med} \), except for the range from which the ratios are drawn, the most
unbiased assumption is to assume a uniform probability \( p \) within this
range. Hence, the probability of finding a log-ratio in \( I_{med} \) is \( p=\frac{1}{l} \)
and \( p'=\frac{1}{L} \) for an outlier. We then retain \( I_{med} \) that
maximizes the likelihood of the full probe set ratios given our model. We must
therefore maximize \( \cal L\;  \)\( =-(N-n)\log L-n\log l \), where \( N \)
is the total number of cells and \( n \) the cells in \( I_{med} \). In essence,
this procedure picks the optimal interval \( I_{med} \) as a tradeoff between
having too many outliers, and letting the range of \( I_{med} \) become too
wide. Prototype situations showing how our model selects \( I_{med} \) are
presented in Fig. 5 for two different conditions. After having identified \( I_{med} \),
we compute scores by taking geometric means of cell ratios and intensities from
PM cells inside \( I_{med} \).

It is now worthwhile looking back at Fig. 1(b) and (c) showing which probe cells
are actually selected. As a fact, there are only few probe sets that have common
low intensity cells(cf. Fig. 2A). Instead, the low intensity cells are distributed
among the probe sets, which is clearly reflected in Fig. 1(c) by the low density
of points at the low end.

\section*{Results}

To demonstrate the potential of our method, we analyzed a set of HDONA hybridizations
evaluating the transcriptional profiles of six different mouse brain regions
using the Mu11k mouse A and B chip series. The dissections and enzymatic steps
(making the target cRNA) were performed in duplicate in all experiments and
the two obtained samples were hybridized onto separate arrays. Fig. 5A shows
the scatter plots of the replicates from four brain regions, the A and B chips
being superimposed on the same plot. 

Our scores exhibit a much tighter scatter, especially in the mid to low intensity
range. Further, we are able to report scores for all the genes on the arrays,
whereas the Affymetrix algorithm reports non-negative values (negative intensities
are meaningless and not plotable on a logarithmic scale) for \( \sim 70\% \)
of the probe sets. As mentioned, the reason we obtain relatively few low intensity
genes (\( 2\sigma  \) of residual background \( \sim 100 \) in these units)
is that low intensity cells tend to be distributed among different probe sets
rather than being grouped. The histograms in Fig. 5B show the distributions
of the \( \log _{2} \) ratios from the four combined regions in intensity windows.
Our distributions are well fitted by narrow Gaussians with standard deviations
\( \sigma \sim 0.2 \) for intensities \( >300 \). \( 2\sigma  \) then corresponds
to a fold change of \( \sim 1.25 \). In contrast, the Affymetrix scores lead
to longer tails especially in the mid to low intensity range. Next, we contrast
a replicate experiment Fig. 6(a) with a comparison of two different experimental
conditions Fig. 6(b). As \( \sigma  \) is not strongly intensity dependant,
we have tentatively indicated in red the fold changes of \( 1.25 \). Consistently,
\( 6\% \) of the genes lie beyond the \( 2\sigma  \) lines in (a). For the
comparison of two different condition (b), \( 20\% \) of the genes are differentially
expressed by a factor \( \geq 1.25 \). We should further mention that the location
of our points in the scatter plot is equivalent to the reported ratio, which
is not the case for the Fold Change calculated by Affymetrix.

Finally, we demonstrate that our procedure leads to a greatly enhanced consistence
between the scores obtained from probe sets for identical genes. There are \( \sim 700 \)
genes represented twice or more on the combined A and B Mu11k mouse chips. The
sequences for two such sets may probe different locations on the same gene,
or one probe set may represent a subsequence of the other. Nevertheless, they
correspond to the same physical gene and should ideally lead to identical scores.
Let \( p_{1} \) and \( p_{2} \) be two such probe sets for a common gene,
and \( r_{i} \) (\( i=1,2 \)) the ratio of the \( p_{i} \) intensities probed
in two different brain regions. In Fig. 7, we show the distributions of \( \log _{2}(\frac{r_{1}}{r_{2}}) \)
for two comparisons C1 and C2 (C2 corresponds to Fig. 1(b)). The figure is separated
into left and right according to whether the Affymetrix fold change was reported
with a ``\( \sim  \)'' in at least one of the two representations \( p_{1} \)
or \( p_{2} \) (the ``\( \sim  \)'' indicates that the baseline intensity
was within the residual background, suggesting that the reported value is unreliable).
It is evident that our ratios are far more consistent than the Affymetrix scores,
especially in the right panels. Our standard deviations \( \sigma  \) are similar
throughout all plots, the \( \sigma  \) on the right panels being barely larger.
Taking \( 2\sigma \sim 0.4 \) implies that \( 95\% \) of the pairs (\( r_{1},r_{2}) \)
differ in ratios by a factor less than \( \sim 1.3 \), which is a significant
narrowing in comparison to the distributions produced by the current Affymetrix
algorithm.

\section*{Summary}

We presented an improved approach for computing composite ratio scores for high-density
oligonucleotide arrays. Our new method differs significantly from the current
Affymetrix algorithm in the following manner: (i) MM cells are not included
because their information content is unclear; (ii) ratios between two different
samples are derived from comparing the PM cell intensities pairwise, and then
identifying a subset of probe cells leading to optimally consistent scores;
(iii) geometric averages are used because the intensity and ratios of probe
sets are distributed on a exponential scale. We showed that our method acts
as a noise reducing filter in the sense that (i) replicate experiments show
an increased reproducibility; (ii) ratio scores for probe sets probing the same
gene show a much greater correlation. We emphasized that because the distribution
of intensities within a probe set is often much broader than the distribution
of cell ratios taken from a pairwise comparison, the most reliable information
lies in ratio and not in absolute intensity composites. Therefore, we designed
our algorithm to primarily compute ratios, and reported intensities (e.g. in
scatterplots) always dependent upon a comparison. 

Our method should be considered as a the simplest way to extract the most information
from just two hybridization arrays, therefore allowing benefits from working
with small data sets of very homogeneous quality. In contrast, the more involved
model-based approach\cite{li} requires large data sets for calibration, and
is therefore more sensitive to the variability introduced by slight changes
in the experimental protocol. Further, knowledge of the probe sequences would
enable one to develop more elaborate approaches based on the kinetic and thermodynamic
properties of the probes.

We have applied our method to a large microarray data set studying the neurogenesis
in adult mice brains, which lead to highly significant biological results\cite{dan}.
The generated data sets could be clustered robustly using standard hierarchical
techniques. Finally, the fact that MM cells are not explicitely needed opens
the possibility of screening twice as many genes on a given microarray. Considering
that the current estimates about the number of genes in the human genome predict
fewer than \( 40,000 \) genes, it is not unrealistic to expect single arrays
for all human genes in the near future.\\

We thank E. van Nimwegen for suggesting many interesting ideas. We benefitted
from inspiring discussions with J. Luban, M. Asmal and A. Alvarez-Buylla. Help
from C. Hacker has been invaluable for generating the extremely high quality
data. F. N. acknowledges the the Swiss National Science Foundation for financial
support. D. A. L. was supported by the NIH grant GM07739.

\end{document}